\documentstyle[aps,pra,epsfig,twocolumn]{revtex}

\def\be{\begin{equation}}
\def\ee{\end{equation}}
\def\bea{\begin{eqnarray}}
\def\eea{\end{eqnarray}}
\def\bma{\begin{mathletters}}
\def\ema{\end{mathletters}}
\def\tr{{\rm tr}}

\def\C{\hbox{$\mit I$\kern-.7em$\mit C$}}
\def\st{\mbox{ s. t. }}
\newcommand{\one}{\mbox{$1 \hspace{-1.0mm}  {\bf l}$}}
\begin{document}
\draft

\title{Characterization of separable
states and entanglement witnesses}

\author{M. Lewenstein$^{1}$, B. Kraus$^{2}$, 
P. Horodecki$^{3}$, and J. I. Cirac$^{2}$}

\address{
$^1$ Institute for Theoretical Physics,
University of Hannover, Hannover, Germany\\
 $^2$ Institute for
Theoretical Physics, University of Innsbruck, A--6020 Innsbruck,
Austria\\
 $^3$ Faculty of Applied Physics and Mathematics,
Technical University of Gda\'nsk, 80--952 Gda\'nsk, Poland}

\date{\today}

\maketitle
\begin{abstract}
We provide a canonical form of mixed states in bipartite quantum
systems in terms of a convex combination of a separable state
and a, so--called, {\it edge} state. We construct entanglement
witnesses for all edge states. We present a canonical form of
nondecomposable entanglement witnesses and the corresponding
positive maps. We provide constructive methods for their
optimization in a finite number of steps. We present a
characterization of separable states using a special class of
entanglement witnesses. Finally, we present a nontrivial
necessary condition for entanglement witnesses and positive maps
to be extremal.
\end{abstract}

\narrowtext

One of the most fundamental open problems of quantum mechanics
is the characterization and classification of mixed entangled
states of multipartite systems, i.e., states that exhibit
quantum correlations \cite{review}. This problem is of
enormous importance for applications in quantum information
processing\cite{Ekert,geste,tel,comput}. A density operator
$\rho\ge 0$ acting on a finite Hilbert space $H=H_A\otimes H_B$
describing the state of two quantum systems $A$ and $B$ is
called entangled \cite{We89} (or not separable) if it {\it cannot} be
written as a convex combination of product states, i.e., as
\be
\label{rhosep}
\rho = \sum_k p_k |e_k,f_k\rangle\langle e_k,f_k|,
\ee
where $p_k\ge 0$, and $|e_k,f_k\rangle\equiv
|e_k\rangle_A\otimes|f_k\rangle_B$ are product vectors.
Conversely, $\rho$ is separable (or not entangled) if it can
be written in the form (\ref{rhosep}).

For low dimensional systems (in $H=\C^2\otimes\C^2$ and
$H=\C^2\otimes\C^3$), there exists an operationally simple
necessary and sufficient condition for separability, the
so--called Peres--Horodecki criterion \cite{Pe96,Ho96}. It
indicates that a state $\rho$ is separable iff its partial
transpose is positive, where partial transpose means the
transpose with respect to one of the subsystems \cite{pt}.
However, in higher dimensions this is only a necessary
condition; that is, there exist entangled states whose partial
transpose is positive (PPTES) \cite{Ho97,Be98,bound}.  Thus, the
separability problem reduces to finding whether a density
operator with positive partial transpose (PPT) is separable or
not \cite{review}.

There exists a complete characterization of separable states
based on entanglement witnesses (EW) and positive maps (PM)
\cite{Ho96}. Briefly speaking: a state $\rho$ is entangled iff
there exists a hermitian operator $W$ (an EW) such that ${\rm
Tr}(W\sigma)\geq 0$ for all separable $\sigma$, but ${\rm
Tr}(W\rho)<0$. The latter condition offers the possibility of
experimental detection of entanglement via the measurement of
$W$ -- an observable which ``witnesses'' the quantum
correlations in $\rho$\cite{Te99}. Starting from EW's one can
define PM's \cite{Ja72} that also detect entanglement. An
example of a PM is transposition, $T$\cite{St63,Wo76}, whose
tensor extension $I\otimes T$ detects all non PPT states.
Unfortunately, the characterization of EW's and PM's is not
known, and therefore the most challenging open questions are:
How to construct EW's in general, and what is the minimal set of
them which allows to detect all entangled states. First steps
toward answering these questions have been accomplished in Ref.
\cite{Te99}.

In this Letter we provide a canonical form of mixed states in
bipartite quantum systems in terms of a convex combination of a
separable state and a so--called {\it edge} state, which
violates extremely the range separability criterion\cite{range}.
We construct EW's for all edge states and present a canonical
form for nondecomposable EW (nd--EW) and the corresponding PM.
We present constructive methods to optimize nd--EW's in a finite
number of steps. We provide a characterization of separable
states using a special class of EW's that are not necessarily
related to edge states, but to certain subspaces of $H$.
Finally, we present a nontrivial necessary condition for
nd--EW's and PM's to be extremal. The methods that we use to
prove our result are based on the technique of ``subtracting
projectors on product vectors''\cite{Le98,Kr99}. Most of the
technical proofs have been included in Ref. \cite{Le00}.

In this paper we will denote by $K(\rho)$, $R(\rho)$, and
$r(\rho)$ the kernel, range, and rank of $\rho$, respectively.
Let us start by defining the edge states. An ``edge'' state,
$\delta$, is a PPTES such that for all product vectors
$|e,f\rangle$ and $\epsilon>0$, $\delta-\epsilon
|e,f\rangle\langle e,f|$ is not positive or does not have a PPT.
Obviously, the ``edge'' states lie on the boundary between PPTES
and not PPT states. In order to characterize them we use the
following criterion \cite{Ho97,Kr99}:

\noindent{\bf Criterion:} A PPT state
$\delta$ is an ``edge'' state iff there exists no
$|e,f\rangle\in R(\delta)$ s.t. $|e,f^\ast\rangle\in
R(\delta^{T_B})$.

Note that the edge states violate the range criterion of
separability in an extreme manner\cite{Ho97,Kr99}. They are of
special importance since they are responsible for the
entanglement contained in PPTES's. In order to see that we
generalize the method of the best separable approximation (BSA)
\cite{Le98} to the case of PPT states:

\noindent{\bf Proposition 1:}
Every PPTES $\rho$ is a convex combination
\be
\label{decomp}
\rho=(1-p) \rho_{sep} + p \delta,
\ee
of some separable state, $\rho_{sep}$, and an edge state, $\delta$.

Note that in the decomposition (\ref{decomp}) the weight $p$ can
be chosen to be minimal [i. e. there exists no decomposition of
type (\ref{decomp}) with a smaller $p$].

The decomposition (\ref{decomp}) can be obtained using the
method of subtracting projectors onto product states
$|e,f\rangle\in R(\rho)$ such that $|e,f^*\rangle\in
R(\rho^{T_B})$. One can show \cite{Le98} that $\rho'\propto
\rho-\lambda |e,f\rangle\langle e,f|$ is still a PPTES
if $\lambda=\min[1/(\langle e,f|\rho^{-1}|e,f\rangle),
1/(\langle e,f^*|(\rho^{T_B})^{-1}|e,f^*\rangle)]$. Moreover,
such operation diminishes either the rank of $\rho$ or
$\rho^{T_B}$, or both. The construction of the optimal
decomposition is a hard task, but construction of a
decomposition with non minimal $p$ can be obtained in a finite
number of steps. This provides us with a simple method to
construct edge states in arbitrary dimensions, and a
separability check \cite{Kr99}.

It is natural to ask how to detect PPTES, in the view of the
decomposition (\ref{decomp}). As mentioned above, one approach
is to use EW. There exists a class of EW (called
decomposable\cite{Le00}) which have the form $W=P+Q^{T_B}$,
where $P$ and $Q$ are positive operators. Such witnesses can
only detect non PPT entangled states \cite{foot}. The EW which
cannot be written as $W=P+Q^{T_B}$ are called nondecomposable
EW. An EW is non--decomposable iff it detects a PPTES
\cite{Le00}. In particular, every nd--EW detects an edge state
since one can immediately see from (\ref{decomp}) that if ${\rm
Tr}(W\rho)<0$ then ${\rm Tr}(W\delta)<0$. Despite their
importance, it is not known how to characterize the class of
nd--EW's. It is thus an important task to study the EW's of the
edge states.

One of the important results of this letter is that for any edge
state one can explicitely construct a nd--EW which detects it.
To show that, we generalize the method of \cite{Te99}, which is
restricted to PPTES constructed out of unextendible product
bases \cite{Be98} which, in particular, do not exist for
$2\times N$ dimensional systems. Let $\delta$ be an edge state,
$C$ an arbitrary positive operator such that ${\rm Tr}(\delta
C)>0$, and $P$ and $Q$ positive operators whose ranges fulfill
$R(P)\subseteq K(\delta)$, $R(Q)\subseteq K(\delta^{T_B})$. We
define
\be
\label{Wd}
W_\delta \equiv P + Q^{T_B},
\ee
and
\be
\label{epsilon1}
\epsilon\equiv \ \inf_{|e,f\rangle}\langle
e,f|W_{\delta}|e,f\rangle; \ \ c\equiv \sup_{|e,f\rangle}\langle
e,f|C|e,f\rangle.
\ee
Note that the properties of $\delta$ ensure that $\epsilon > 0$.
We then have

\noindent{\bf Lemma 1:} [Lemma 6 of Ref. \cite{Le00}]
Given an edge state $\delta$, then
\be
W_1 = W_\delta-{\epsilon \over
c}C
\ee
is a nd--EW which detects $\delta$.

The simplest choice of $P$, $Q$ and $C$ consists of taking the
projections onto $K(\delta)$, $K(\delta^{T_B})$ and the identity
operator, respectively \cite{choice}. As we will see below, this
choice provides us with a canonical form for nd-EW. In order to
show that, let us first introduce some additional notations.

Let $\cal S \subset {\cal P}$ denote the convex set (cone) of
separable (resp. PPT) states. Let ${\cal P}^{\perp} \subset
{\cal S}^{\perp}$ be the convex sets (dual cones) of nd-EW's
(resp. EW's). All those sets are closed.

\noindent{\bf Definition:}
An EW (resp. decomposable EW), $W$ is {\it tangent} to ${\cal
S}$ (resp. {\it tangent} to ${\cal P}$) if there exists a state
$\rho \in {\cal S}$ ($\rho \in {\cal P}$) such that ${\rm
Tr}(W\rho)=0$. Furthermore, we say that $W$ is tangent to ${\cal
S}$ (${\cal P}$) at $\rho$ $\in{\cal S}$ (${\cal P}$) if ${\rm
Tr}(W\rho)=0$.

\noindent{\bf Observation 1:} The state $\rho$ is separable iff
for all EW's tangent to ${\cal S}$, ${\rm Tr}(W\rho)\ge 0$.

\noindent{\it Proof:} (only if) is trivial; (if)
Let $\rho$ be an entangled state, and let $W$ be an EW that
detects $\rho$, i.e., ${\rm Tr}(W\rho)<0$. We define
$\epsilon\ge 0$ as in (\ref{epsilon1}). If $\epsilon=0$ then $W$
is tangent to ${\cal S}$. If $\epsilon>0$ then $W'=W-\epsilon
\one$ is still an EW which detects $\rho$ and it is tangent to
${\cal S}$.

\noindent{\bf Observation 2:}
If a decomposable EW, $W$, is tangent to $\cal P$ at $\rho$,
then for any decomposition (\ref{decomp}) $W$ must also be
tangent to $\cal P$ at the edge state $\delta$.

We can prove now

\noindent{\bf Proposition 2:} If an EW, $W$, which does not detect any
PPTES, is
tangent to  $\cal P$ at some edge state $\delta$ then it is of the form
\bea
W=P+Q^{T_{B}}
\eea
where $P, Q\geq 0$ such that $R(P)\subseteq K(\delta)$, $R(Q)\subseteq
K(\delta^{T_B})$.

\noindent {\it Proof:} As mentioned before, an EW, $W$, which does not
detect
any PPTES must be decomposable; that is, $W=P+Q^{T_{B}}$. From the PPT
property
of $\delta$ and the positivity of $P, Q$ we have that the ranges $R(\delta)$
and
$R(P)$ [resp. $R(\delta^{T_B})$ and $R(Q)$] must be orthogonal.

We are now in the position to prove one of the main results of this paper,
regarding our canonical form of nd--EW's:

\noindent{\bf Proposition 3:} Any nd--EW, $W$,
has the form
\bea
W=P+Q^{T_{B}} - \epsilon \one, \ \ 0< \epsilon\le \
\inf_{|e,f\rangle}\langle e,f|P +Q^{T_{B}}|e,f\rangle.
\label{complet}
\eea
where $P$ and $Q$ fulfill the conditions of Prop. 2 for some edge state
$\delta$.

\noindent{\it Proof:} Consider $W(\lambda)=W+\lambda\one$. Obviously for
some
$\lambda>0$, say $\lambda_0$, $W(\lambda_0)$ becomes decomposable. Note that
for
any $\lambda<\lambda_0$, $W(\lambda)$ is nondecomposable and therefore it
detects some PPTES $\rho$. Using continuity we conclude that $W(\lambda_0)$
is
tangent to $\cal P$. From Obs. 2 there exists an edge state $\delta$ to
which
$W(\lambda_0)$ is tangent. From Prop. 2 we obtain that $W(\lambda_0)=P
+Q^{T_{B}}$, where $P$ and $Q$ satisfy the needed conditions, and
consequently
$W=P +Q^{T_{B}}-\epsilon\one$ with $\epsilon=\lambda_0$. Since $W$ is EW,
$\epsilon$ must not be greater than $\inf_{|e,f\rangle}\langle e,f|P
+Q^{T_{B}}|e,f\rangle$.

\noindent{\bf Proposition 3':} If the assumptions of Prop. 3 hold then $W$
is of
the form (\ref{complet}) with $R(P)$, $R(Q)$ orthogonal to some Hilbert
subspaces ${\cal H}^a$ and ${\cal H}^b$, respectively, where: (i) there
exists
no $|e,f\rangle \in{\cal H}^a$ s.t. $|e,f^*\rangle \in{\cal H}^b$; (ii)
$R[{\rm
Tr}_B (P_{{\cal H}^a})]=R[{\rm Tr}_B (P_{{\cal H}^b})]$, $R[{\rm Tr}_A
(P_{{\cal
H}^a})]=R[{\rm Tr}_A (P_{{\cal H}^b})^*]$, where $P_{X}$ stands for the
projector onto the subspace $X$; (iii) ${\rm dim}{\cal H}^x>\max [r({\rm
Tr}_A
(P_{{\cal H}^x})), r({\rm Tr}_B (P_{{\cal H}^x}))]$, $x=a,b$.

\noindent{\it Proof:} The point (i) is clear; (ii) and (iii) follow from
simple
analysis of the ranges of the partial reductions of $\delta$ as well as the
properties of the range of PPT states \cite{Ho0a,Kr99}.

\noindent{\bf Remark 1:} The presented formulation permits to release
ourselves
from dealing with edge states in the canonical decomposition
(\ref{complet}).
Instead, we may consider only the pairs of ``strange'' subspaces $H^{a,b}$
of
the Hilbert space. Note that the converse of the Proposition 3' is also
likely
to be true.

\noindent{\bf Remark 2:} It is worth recalling  that all EW's are in one to
one
correspondence to PM's\cite{Ja72}. In particular, any nd-EW leads to a
so--called {\it nondecomposable positive map } (nd-PM), i.e., a map which
cannot
be written as a convex sum of a completely positive map and some other
completely positive map followed by transposition. The characterization of
nd-PM's is one of the most challenging open problems in mathematical physics.
Prop. 3 (3') thus provides us with a {\it canonical form for nd--PM}. As we
mentioned, a PM $\Lambda$ (transforming operators acting
on
${\cal H}_C$ to those acting on ${\cal H}_B$) provides a separability test
which
is stronger than its EW counterpart $W_{\Lambda}$ acting on ${\cal
H}_A\otimes{\cal H}_B$. The correspondence between such PM and EW is given
by
the following relation: if $|\Psi\rangle=\sum_{k=1}^{d_A}|k\rangle_A\otimes
|k\rangle_C$ then $W_{\Lambda}=\one_A\otimes\Lambda(|\Psi\rangle \langle \Psi|) $.

As mentioned above, when studying separability we just have to deal with
nd--EW's. In order to reduce the set of nd--EW's and nd--PM's,
let us introduce
the following definitions. Given two nd--EW's, $W_1$ and $W_2$, then we say
that
$W_2$ is {\it nd--finer} than $W_1$ if all the PPTES detected by $W_1$ are
also
detected by $W_2$. We say that $W$ is a nondecomposable {\it optimal}
(nd--optimal) EW (nd--OEW), if there exists no nd--EW which is nd--finer
than
$W$. Thus it is obvious that the nd--EW's we are interested in are the
nd--OEW's. Let us  call
an
operator $D=P+Q^{T}$, with $P,Q\geq 0$ and $T$ denoting the 
partial transposition with respect to  $A$ or $B$, 
decomposable. Furthermore let us define the set of
product vectors on which the average of
$W$ vanishes, i.e., $p_W=\{|e,f\rangle\in H, \st
\langle e,f|W|e,f\rangle=0\}$. This set plays an important
role in the optimization, which can be seen in the following results
concerning
the characterization of nd--OEW's:

\noindent{\bf Proposition 4:} [Theorem 1b of Ref. \cite{Le00}] An nd--EW,
$W$,
is nd--optimal iff for all decomposable operators $D$ and $\epsilon > 0$ the
operator $W'= W - \epsilon D$ is not an EW.

\noindent{\bf Corollary :} If both $p_W$ and $p_{W^T}$ span the whole Hilbert
space,
$H_{A} \otimes H_{B}$, then $W$ is a nd--OEW.

\noindent{\bf Remark 3:} The necessary and sufficient conditions for a
nd--EW to
be nd--optimal are presented in Ref. \cite{Le00}. Loosely speaking a nd--EW
is
nd--optimal iff either both, $p_W$ and $p_{W^T}$ span the whole Hilbert space, or
there exist some nonproduct vectors $|\Psi\rangle$ related to $p_W$ (or
$p_{W^T}$) s.t. both, $p_W$ ($p_{W^T}$) joint with the set of $|\Psi\rangle$ 
and $p_{W^T}$ ($p_W$) span the whole
$H_{A} \otimes H_{B}$. In our numerical studies, however, we have not
encountered  the latter possibility; it is thus likely that the converse of
the
Corollary is true.

Our results allow now to design a finite step algorithm to nd--optimize a
given
EW, $W$, by subtracting decomposable operators:

\noindent(I) Take a decomposable operator, $D=P+Q^T$ such that $Pp_W=0$ and
$Q p_{W^T}=0$ and check if
\be
\label{lambdap}
\lambda_0\equiv
\inf_{|e\rangle\in H_A} \left[D_e^{-1/2} W_e D_e^{-1/2}\right]_{\rm min} \\
>0.
\ee
Here $W_e=\langle e|W|e\rangle$, $D_e=\langle e|D|e\rangle$,  where
$|e\rangle\in H_A$, whereas  $[X]_{\rm min}$ is the minimal eigenvalue of
$X$.

\noindent(II) If $\lambda_0$ is positive construct the new, nd--finer
EW, $W'=W-\lambda_0 D$.

\noindent(III) Iterate the procedure (I)-(II) as long as there is no
$D=P+Q^T$ with $P p_W=0$ and $Q p_{W^T}=0$.

After each step the set of $p_W$, $p_{W^T}$ or both of them increases at
least
by one element, which is linearly independent of the former ones. So, after
a
finite number of steps $p_W$ and $p_{W^T}$ will span the whole Hilbert space
which ensures that the final nd--EW is nd--optimal. In principle, it may
happen
that $\lambda_0=0$ at some step, before $p_W$ and $p_{W^T}$ span the whole
$H$. Our numerical simulations suggest, however, 
 that among all possible $D$'s one can always find one 
 with $\lambda_0>0$.

We have applied the methods of finding and optimizing EW's to a family
$\rho_b$
($b \in [0, 1]$) of PPTES  in the $2 \times 4$ dimensional system from Ref.
\cite{Ho97}. For $b=0,1$ those states are separable, whereas for $0<b<1$,
$\rho_b$'s are edge states which can be checked directly as shown in Ref.
\cite{Ho97}. We have applied the following procedure. By virtue of some
symmetries of $\rho_b$, one can perform a local change of basis after which
the
transformed state $\tilde{\rho_b}$  fulfills
$\tilde{\rho}_b^{T_B}=\tilde\rho_b$.
This step allowed us to construct the nd--EW
$W_1=P+P^{T_B}-\lambda_0\one$,
which detects already the edge state. Following the procedure above we
subtracted decomposable operators. In addition we choose them to be
invariant under partial
transposition with respect to system B. Note that then $W=W^{T_B}$ at any
step
and therefore we only had to make sure that $p_W$ spanned the whole
Hilbert space,
which automatically ensured that the final nd--EW was nd--optimal. In Fig. 1
it
is shown how many members of the whole family of $\rho_{b'}$'s are detected
by
the nd--OEW obtained from $\rho_b$. We plot here also the efficiency of the
corresponding nd-PM. Here the improvement of efficency is less spectacular,
but
still significant.

It must be stressed that both: the EW's and the PM's constructed in $2\times
4$
system are the first examples provided for a quantum system with one qubit
subsystem. We have also provided the first examples of the set $p_W$ that
spans
the whole Hilbert space. This set allows to construct very peculiar
separable
states of full rank that lie on the boundary of $\cal S$. Note also that, in
general, the parameter $\lambda_0$ in the optimization procedure has to be
found
numerically. In Ref. \cite{Le00} we have been able to formulate an analytic
method that allows to detect the whole family of $\rho_b$'s.

\begin{figure}[ht]
\begin{picture}(200,200)
\put(5,5){\epsfxsize=200pt\epsffile[23 146 546 590 ]{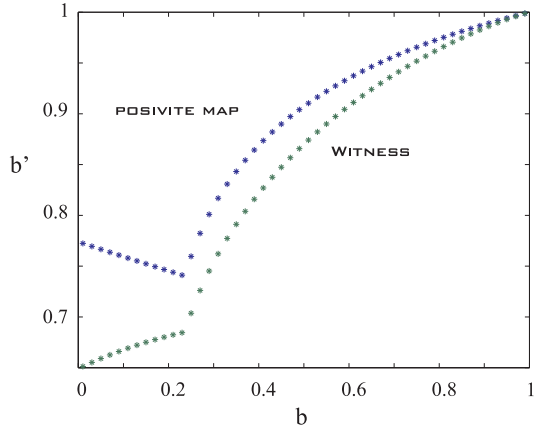}}
\end{picture}
\caption[]{Values of $b'$ for which if $\tilde b\le b'$,
$\tilde\rho_{\tilde b}$
is detected by the optimal witness and the positive map obtained from
$\tilde\rho_b$.}
\label{Fig1}
\end{figure}

As we remember, the key problem is to find the minimal set of EW's detecting
PPTES. Obviously, this minimal set will consist of nd-OEW's. A related
problem
is to find a set of extremal points of ${\cal P}^{\perp}$. Note that a
nonoptimal nd-EW is a convex sum of an optimal one and a decomposable
operator
(Prop.4), so it cannot be an extremal point. Note that Prop. 3 (3') combined
with the optimality property provides {\it the necessary form of extremal
points} of both EW's, as well as PM's, which has not been known so far. We
have
thus the following

\noindent{\bf Proposition 5:} The set of extremal points of the set of EW's,
${\cal S^\perp}$, is contained in the set $\cal A$ of all optimal EW's of the form
(\ref{complet}) plus projectors and transposed projectors.

\noindent{\bf Proposition 5':} The set of extremal points of the cone of
nd--EW's, ${\cal P^\perp}$, is contained in the set $\cal B$ of all optimal nd-EW's
of
the form (\ref{complet}).

\noindent{\bf Remark 4:} Moreover, applying the isomorphism \cite{Ja72} to
the
members of $\cal A$ ($\cal B$) we obtain the set $\cal A'$ ($\cal B'$) of
PM's
(nd-PM's) containing the  set of all extreme PM's. The above theorems
provide
thus the first nontrivial necessary condition  for EW's and PM's to be
extremal.
In particular, following Prop. 3' we can obtain a weaker condition by
considering optimal EW's of the form (\ref{complet}) without involving the
notion of the edge states,  but only pairs of ``strange'' subspaces ${\cal
H}^a$
and ${\cal H}^b$.

 This work has been supported  by the DFG (SFB 407 and Schwerpunkt
"Quanteninformationsverarbeitung"), the DAAD, the \"OFW  (SFB ``Control and
Measurement of Coherent Quantum Systems''), the ESF PESC Programm on Quantum
Information, TMR network ERB--FMRX--CT96--0087, the IST Programme EQUIP, and
the
Institute for Quantum Information GmbH.

%=========================


\begin{references}
\bibitem{pt}
Given an operator $X$ and an orthonormal basis
$\{|k\rangle\}_{k=1}^N\in H_B$, one defines the partial
transposed of $X$ with respect to $B$ in that basis as follows:
$
X^{T_B} = \sum_{k,k'=1}^N |k'\rangle_B\langle
k|X|k'\rangle_B\langle k|$.
One can analogously define the partial transposed of $X$ with
respect to $A$ in a given basis, $X^{T_A}$. Partial
transposition fulfills the following useful property
$
\label{PropPT}
\tr(X^{T_B} Y)=\tr(XY^{T_B})
$.
We say that positive operator has a positive partial
transposition (PPT) if $\rho^{T_B}\ge 0$. Note that this
property is basis independent.
\bibitem{range}According to this criterium, if $\rho$
is separable then there must
exist a set of product vectors $\{|e_k,f_k\rangle\}$ that spans
the range $R(\rho)$ s. t. the set of partial complex
conjugated product states $\{|e_k,f^\ast_k\rangle\}$ spans the
range of $\rho^{T_B}$, see \cite{Ho97}.
\bibitem{foot}
Optimal decomposable EW can be shown to have the form of the
partial transpose of a positive operator whose range does not
contain any product vector; see \cite{Le00}.
\end{references}
\end{document}